\documentclass{neutrinos19}

\usepackage{physics}
\usepackage{amsmath}
\usepackage{amssymb}  
\usepackage{graphicx}  
\usepackage{verbatim} 
\usepackage{color}    
\usepackage{subfigure}  
\usepackage{hyperref}  
\usepackage{mathrsfs}
\usepackage{float}

\def\be{\begin{equation}}
\def\ee{\end{equation}}
\def\bea{\begin{eqnarray}}
\def\eea{\end{eqnarray}}

\newcommand{\Photo}{}

\begin{document}
\vspace*{4cm}
\title{The 2-Neutrino-Exchange Potential with Mixing:\\A New Arena for Neutrino Mixing 
	and CP-Violation }
\author{ Q.\ Le Thien$^{1}$ and D.E.\ Krause$^{1,2}$}
\address{$^1$Physics Department, Wabash College, Crawfordsville, IN 47933, USA \\ 
	$^2$Department of Physics and Astronomy, Purdue University, \\ 
	West Lafayette, IN 47907, USA}
\maketitle
\abstracts{The 2-neutrino exchange potential (2NEP) is a Standard Model (SM) weak potential due to the exchange of virtual neutrino-antineutrino pairs. Consequently, many aspects of neutrino physics, such as the number of flavors, their masses, fermionic nature (Dirac or Majorana), low-energy neutrino physics and CP-violation, can be examined via the 2NEP. We present a new approach for calculating the 2NEP taking into account the phenomenon of neutrino mixing and CP-violation which arises from the structure of the SM weak interaction Lagrangian. Lastly, we explore implications of our result in various physical contexts. }

\section{Introduction}

Ever since the birth of the weak interaction, neutrinos have been under intense study in particle physics due to the strong possibility of their connection to Beyond Standard Model (BSM) physics. Yet, even decades after their first detection and the surprising discovery of neutrino mixing, many fundamental physical properties of neutrinos are still currently not sufficiently elucidated, i.e. their exact masses, their fermionic nature (Dirac or Majorana), CP-violation and their physics at low-energy. While certainly tremendous efforts, both theoretical and experimental, have been focused on studying these aspects of the neutrinos, it is important to recognize that in current neutrino experiments we only have access to ultrarelativistic neutrinos. Given the smallness of the neutrino mass, the experimental limitation to ultrarelativistic neutrinos presents a serious obstacle. In this work, we will discuss a phenomenon which involves contributions from all neutrinos' properties at all energies: the 2-Neutrino Exchange Potential (2NEP). While the magnitude of this force is small, which is the main difficulty for observing it, we hope that our and others' recent works \cite{Stadnik,Asaka,LeThien} will inspire a new arena to study fundamental properties of neutrinos. 

Historically, the 2NEP has long been of theoretical interest. Feinberg and Sucher \cite{feinberg} were the first to derive the famous $1/r^5$ potential by applying dispersion-theoretic techniques to the effective low-energy four-fermion interaction; assuming massless neutrinos, they obtained 
\begin{equation}
V_{\nu,\bar{\nu}} = \frac{G_F^2}{4\pi^3 r^5}.
\end{equation}
In 1995, Fischbach et al. \cite{fischbach et al} used the Hartle-Schwinger formula to calculate the weak interaction contribution, whose dominant term is the 2NEP, to the nuclear binding energy for tests of weak equivalence principle. Under this formalism, Fischbach obtained the important formula for the 2NEP with massive neutrinos \cite{Fischbach AoP}
\begin{equation}
V_{\nu,\bar{\nu}} = \frac{G_F^2 m_\nu^3}{4\pi^3 r^2} K_3\left(2 m_\nu r\right).
\label{2NEP massive 1 flavor}
\end{equation} 
In 1996, applying this result to the problem of the weak binding energy in neutron stars, he was able to put forth the constraint for the lightest neutrino mass \cite{Fischbach AoP}
\begin{equation}
\label{fischbach mass limit}
m_{\nu, \text{ lightest}}\gtrsim 0.4 \text{ eV}.
\end{equation}
Even though this limit is currently inconsistent with constraints from cosmology ($m_{\text{tot}} < 0.152 \text{ eV}$),\cite{choudhury}  it is important to realize that in arriving at his limit, Fischbach did not account for the phenomenon of neutrino mixing, which had not been established until 1998, yet certainly it can significantly affect the result in Eq.~(\ref{fischbach mass limit}).  Additionally, this tension can also stem from the fact that astrophysical limits are model-dependent. Therefore, it is clear that there is a motivation to study effects of neutrino mixing on the 2NEP.

\section{Derivation of the Single-Flavor 2-Neutrino Exchange Potential}
\label{derivation section}

As briefly aforementioned, there were previously two main approaches to calculate the 2NEP: dispersion theory \cite{feinberg} and Schwinger formula.\cite{Fischbach AoP} However, these two methods do not explicitly show the dependence of the 2NEP on the structure of the neutrino vacuum, which is found to be non-trivial in many quantum field-theoretic treatments of neutrino mixing.\cite{BV AoP,Tureanu}  Hence, in our approach we decided to identify the 2NEP as a part of the vacuum energy contained in the neutrino fields, rather than to use the modern covariant diagrammatic approach. In this section, we will quickly discuss our approach, assuming there is only one neutrino flavor participating in the exchange for the sake of clarity. 

Because potential energy is an inherently nonrelativistic concept, it is fair for us to assume that the two external particles participating in this interaction are located at ``fixed" (moving at negligible velocities) positions in space, i.e. $\vec{r}_1$ and $\vec{r}_2$. There must be then a change in the vacuum energy of the neutrino field due to the interaction with the two external particles. This interaction is described by the following interaction Hamiltonian
\begin{equation}
H_{\rm int} = \frac{G_{F}}{\sqrt{2}} \int d^{3}r \,J^{f}_{\mu}(\vec{r})\,\left[\bar{\nu}(\vec{r})\gamma^{\mu}\left(1 - \gamma^{5}\right)\nu(\vec{r})\right],
\label{H int general}
\end{equation}   
which has a mathematical form similar to Fermi's four-fermion interaction in the SM. Because we only consider the spin-independent portion of the 2NEP, the fermion current $J^{f}_{\mu}(\vec{r})$ is given by
\begin{equation}
J_{\mu}^{f}(\vec{r}) = J_{0}^{f}(\vec{r})\delta_{\mu,0} =  \left[\delta^{3}(\vec{r} - \vec{r}_{1}) +  \delta^{3}(\vec{r} - \vec{r}_{2})\right]\delta_{\mu,0}.
\label{rho}
\end{equation}
To proceed, we need to isolate out the finite portion from the vacuum energy of the neutrino field in order to exclude the infinite self-energy terms. This task can be done most straightforwardly in the non-covariant Schr\"odinger picture using time-independent Raleigh-Schr\"odinger perturbation theory. In this framework, we can then work explicitly in coordinate space, which helps to identify the potential energy with terms that only depend on the separation distance $r = |\vec{r}_1 - \vec{r}_2 |$ between the two external particles. The result of our approach can be written concisely as \cite{LeThien}
\begin{equation}
E_{\rm vac}^{(2)}(\vec{r}_{1}-\vec{r}_{2}) = 
-\sum_{E^{(0)}_{n}\neq 0}\left[\frac{\langle 0|H_{{\rm int}}\left(\vec{r}_1\right)|E^{(0)}_{n}\rangle \langle E^{(0)}_{n}|H_{{\rm int}} \left(\vec{r}_2\right)|0\rangle}{E^{(0)}_{n}} + {\rm c.c.}\right],
\label{E vac potential}
\end{equation}
where “c.c.” means complex conjugate, which in this case simply interchanges particles at $\vec{r}_1$ and $\vec{r}_2$. The integral can be carried out and shown to reproduce the exact result in Eq.~(\ref{2NEP massive 1 flavor}) obtained by other modern approaches. More importantly, Eq.~(\ref{E vac potential}) shows that the 2NEP explicitly depends on the action of the neutrino field operators $\nu\left(\vec{r}\right)$ on the neutrino free vacuum $|0\rangle $.

\section{Derivation of the 2-Neutrino Exchange Potential with Mixing and CP-Violation}

\subsection{Neutrino Vacuum}

In the standard quantum-mechanical treatment of neutrino mixing for neutrino oscillation scenarios, the neutrino flavor states $|\nu_{\alpha}\rangle$ are related to the neutrino mass states $ |\nu_{a}\rangle$ via the Pontecorvo-Maki-Nakagawa-Sakata (PMNS) matrix $U_{\rm PMNS}$ as follows
\begin{equation}
|\nu_{\alpha}\rangle = \sum_{a=1}^{3} U^*_{\alpha a}|\nu_{a}\rangle.
\label{QM state mixing}
\end{equation}   
However, obtaining the 2NEP is essentially a quantum field-theoretic problem, thus the mixing prescription in Eq.~(\ref{QM state mixing}) needs to be generalized. Indeed, this is a non-trivial task in quantum field theory since aside from the neutrino flavor field operators $\nu_{\alpha} \left(\vec{r}\right)$, which are quite analogous to quantum states, we also need to prescribe how the mixing affects the neutrino free vacuum $|0\rangle$. In this work, we will follow previous works \cite{Shrock PLB,Shrock PRD,Ho} to assume that the mixing of neutrino mass fields leaves the neutrino vacuum invariant. Hence, the neutrino vacuum is straightforwardly the product of the vacuum states of each neutrino mass field,
\begin{equation}
|0\rangle \rightarrow \prod_{a=1}^{3}|0\rangle_{a}.
\label{neutrino vacuum}
\end{equation}    
Hereafter, we will succinctly refer to this vacuum as $|0\rangle$. 

There are two important explications following the vacuum prescription in Eq.~(\ref{neutrino vacuum}). Firstly, the neutrino free vacuum is the eigenstate of the free Hamiltonian, thus in assuming the trivial vacuum structure as in Eq.~(\ref{neutrino vacuum}), we are fundamentally taking the free Lagrangian of the neutrino mass fields to be that of free Dirac fermions. While this is a reasonable assumption used widely in past literature,\cite{Shrock PLB,Shrock PRD,Ho}  it has been pointed out by others \cite{BV AoP,Tureanu} that the phenomenology of neutrino mixing in Eq.~(\ref{QM state mixing}) can also manifest as additional terms in free Lagrangian, ultimately resulting in a non-trivial neutrino free vacuum which is different from Eq.~(\ref{neutrino vacuum}). Even though those additional terms in the free Lagrangian will modify the oscillating formula, in current oscillation experiments with ultrarelativistic neutrinos, these differences will be completely suppressed. On the other hand, the 2NEP, as shown in Section \ref{derivation section}, is particularly sensitive to the non-trivial structure of the neutrino free vacua. Secondly, it can be shown that the following results obtained for the 2NEP with mixing will still persist, even when other neutrino vacua are used. Furthermore, it is clear that since we have assumed a trivial action of neutrino mixing on the neutrino vacuum, the effect of neutrino mixing that we see later in the 2NEP comes purely from the bilinear structure of the SM weak interaction in the low-energy limit.

\subsection{Neutrino Flavor Fields and Interaction Hamiltonian with Mixing}

Because  our approach relies on the mass fields $\nu_a$ while both the Neutral Current (NC) and Charged Current (CC) interaction in the SM weak interaction are expressed in terms of neutrino flavor fields $\nu_{\alpha}$, we firstly need to have a relation between the neutrino flavor fields and neutrino mass fields. From the analogy between quantum states in quantum mechanics and field operators in quantum field theory, we can generalize from Eq.~(\ref{QM state mixing}) this relation as
\begin{equation}
\nu_{\alpha}(\vec{r}) \equiv \sum_{a = 1}^{3}U_{\alpha a} \, \nu_{a}(\vec{r}).
\label{nu alpha a relation}
\end{equation}

It is now forthright for us to consider the interaction Hamiltonians appropriate for our problem. Since a potential is a nonrelativistic concept, we are only interested in interactions between nucleons (protons and neutrons) and charged leptons. For nucleons, the interaction Hamiltonian includes only the NC interaction, which couples universally to all neutrino flavor fields
\begin{equation}
H_{\rm int, N}(\vec{r}_{i}) = H_{\rm int, N}^{\rm NC}(\vec{r}_{i}) =\frac{G_{F}g_{V}^{N}}{\sqrt{2}}\left[\sum_{\alpha = e,\mu,\tau} \nu_{\alpha}^{\dag}(\vec{r}_{i})\left(1 - \gamma^{5}\right)\nu_{\alpha}(\vec{r}_{i})\right],
\label{H N flavor}
\end{equation} 
where N = p, n (protons, neutrons) and $g_{V}^{N}$ is the appropriate NC coupling weak charge. For charged leptons, the interaction Hamiltonian includes both the NC and CC interaction
\begin{eqnarray}
\nonumber
H_{\rm int, \alpha}(\vec{r}_{i}) &=& H_{\rm int, \alpha}^{\rm NC}(\vec{r}_{i}) + H_{\rm int, \alpha}^{\rm CC}(\vec{r}_{i}) \\
&=& \frac{G_{F}}{\sqrt{2}}\left\{ g_{V}^{\alpha}\left[\sum_{\beta = e,\mu,\tau} \nu_{\beta}^{\dag}(\vec{r}_{i})\left(1 - \gamma^{5}\right)\nu_{\beta}(\vec{r}_{i})\right] + \nu_{\alpha}^{\dag}(\vec{r}_{i})\left(1 - \gamma^{5}\right)\nu_{\alpha}(\vec{r}_{i}) \right\}.
\label{H lepton flavor}
\end{eqnarray}
where $\alpha = e, \mu , \tau$ (electrons, muons, taus) and $g_V^\alpha$ is the corresponding NC coupling weak charge. To carry out the calculation, one needs to firstly make use of Eq.~(\ref{nu alpha a relation}) in order to express Eq.~(\ref{H N flavor}) and (\ref{H lepton flavor}) in terms of the mass fields, then substitute the desired particles' interaction Hamiltonians into Eq.~(\ref{E vac potential}) to obtain the 2NEP.

\section{Interaction Potential with Mixing}
\label{result}

\subsection{Potentials for Nucleon-Nucleon Interaction}

Generally, because nucleons interact only through the NC interaction which does not favor any neutrino flavor, the 2NEP between nucleons exhibits no dependence on the mixing parameters. Particularly, the interaction potential is simply the sum of potentials arising from interaction with each neutrino mass field \cite{LeThien}
\begin{equation}
V_{\rm N_{1},N_{2}}(r) = \frac{G_{F}^{2}g_{V,1}^{{\rm N}_{1}}g_{V,2}^{{\rm N}_{2}}}{4\pi^{3}r^{2}}\sum_{a = 1}^{3}m_{a}^{3}K_{3}(2m_{a}r).
\label{general nucleon V}
\end{equation}   
Nonetheless, there are still exponential cut-offs at 3 different length scales corresponding to 3 neutrino masses.

\subsection{Potentials for Nucleon-Lepton Interaction}
\label{nucleon lepton subsection}

Similar to the previous case between nucleons, the 2NEP between a nucleon and a charged lepton also consists of 3 single 2NEPs of each mass field. However, since there is an additional CC interaction at the charged lepton vertex which enhances the interaction with the corresponding neutrino flavor, the nucleon-lepton 2NEP possess dependences on both mixing parameters and the neutrino mass \cite{LeThien}
\begin{equation}
V_{{\rm N}\alpha}(r) = \frac{G_{F}^{2}g_{V}^{\rm N}}{4\pi^{3}r^{2}}\sum_{a = 1}^{3}m_{a}^{3} \left(g_{V}^{\alpha} +|U_{\alpha a}|^{2} \right) K_{3}(2m_{a}r).
\label{general nucleon lepton V}
\end{equation}  
At short distances, when $r \ll m_a^{-1}$ for all neutrino masses $m_a$, the nucleon-lepton 2NEP takes the form \cite{LeThien}
\begin{equation}
V_{\rm N\alpha}(r) \simeq \frac{G_{F}^{2}g_{V}^{\rm N}}{4\pi^{3}r^{5}} \left(3g_{V}^{e} + \sum_{a = 1}^{3}|U_{\alpha a}|^{2}\right) = \frac{G_{F}^{2}g_{V}^{\rm N}}{4\pi^{3}r^{5}} \left(3g_{V}^{e} + 1\right).
\label{massless nucleon electron V}
\end{equation}
Surprisingly, according to Eq.~(\ref{massless nucleon electron V}), the nucleon-lepton 2NEP's dependence on the mixing parameters drops out at short distances. This result can be understood as follows. Through the NC interaction, the nucleon emits all neutrino flavors equally. Meanwhile, at short distances (high momentum-transfer regime), these flavor fields coincide with the mass fields and thus do not exhibit any oscillation. Hence, when they reach the charged lepton, the initial neutrino flavor fields from the nucleon interact with the charged lepton exactly as they were at the nucleon. Therefore, because no oscillation was involved in the process, there should not be any dependence on the mixing parameters.

\subsection{Potentials for Lepton-Lepton Interaction}

The lepton-lepton 2NEP evinces the richest behaviors with respect to mixing since both the NC and CC interaction are involved at both vertices. Firstly, there are two main contributions,
\begin{equation}
V_{\alpha \beta}(r) = V_{\alpha \beta}^{aa}(r) + V_{\alpha \beta}^{a \neq b}(r),
\label{E vac 2 lepton lepton massive}
\end{equation}
where the first term $V_{\alpha \beta}^{aa}(r)$ comes from exchanging neutrino-antineutrino pair of the same mass field, which thus has the same functional form as in the 2 previous cases \cite{LeThien}
\begin{equation}
V_{\alpha \beta}^{aa}(r) =  \frac{G_{F}^{2}}{4\pi^{3}r^{2}}
\sum_{a = 1}^{3} \left[m_{a}^{3} (g_{V}^{\alpha} +|U_{\alpha a}|^{2} ) (g_{V}^{\beta} +|U_{\beta a}|^{2} ) K_{3}(2m_{a}r) \right].
\label{E vac 2 lepton lepton massive aa}
\end{equation}
However, there now also exist terms $V_{\alpha \beta}^{a \neq b}(r)$ arising from exchanging neutrino-antineutrino pairs of different mass fields. Because the integrals corresponding to these terms do not have an exact analytic expression, we expand them to $\mathcal{O} \left[ \left(m_-^{ab}/m_+^{ab}\right)^2 \right]$ 
\begin{equation}
\begin{split}
V_{\rm {\alpha \beta}}^{a \neq b}(r) =    \frac{G_{F}^{2}}{4\pi^{3}r^{2}}
&  \sum_{a > b}^{3}  \frac{ \Re( U_{\alpha a}^* U_{\alpha b}^{}  U_{\beta b}^* U_{\beta a}^{} )  }{4}
\left\{
m_+^{ab} \left[\left(m_+^{ab}\right)^2 + \left(m_-^{ab}\right)^2 \right] K_3\left(\left.m_+^{ab}\right. r\right)
\right. \\
&
\left. \mbox{}
- \frac{4 \left(m_-^{ab}\right)^2 }{r} K_2\left(m_+^{ab}\, r\right) + \mathcal{O} \left[ \left(\frac{m_-^{ab}}{m_+^{ab}} \right)^2 \right] 
\right\},
\end{split}
\label{E vac 2 lepton lepton mixing massive}
\end{equation} 
where $m_{\pm}^{ab} = m_a \pm m_b$. However, this expansion scheme fails when the lightest mass state is actually massless. In this case, the integral can actually be evaluated exactly as \cite{LeThien}
\begin{eqnarray}
\nonumber
V_{\rm {\alpha \beta, m_a=0}}^{a \neq b}(r) &=&   \sum_{\substack{b =1 \\ b \neq a}}^{3}
\frac{ G_{F}^2 \Re( U_{\alpha a}^* U_{\alpha b}^{}  U_{\beta b}^* U_{\beta a}^{} )  } {48 \pi^3 r^5} \left[ 
e^{-m_b r} \left( 24 + 24 m_b r + 6 m_b^2 r^2 -2  m_b^3 r^3 +  m_b^4 r^4 \right. \right.\\
& & \left. \left. - m_b^5 r^5 \right) -  (6 m_b^4 r^4 + m_b^6 r^6) \ \mathrm{Ei}\left(-m_b r\right) - 6 m_b^4 r^4 \ \Gamma\left( 0, m_b r \right) \right].  
\label{E vac 2 lepton lepton massless}
\end{eqnarray}
Distinctively from the aforementioned situations, the lepton-lepton 2NEP has 6 different cut-off length scales due to the additional mixing exchanges from $V_{\alpha\beta}^{a\neq b}$. In the short-range limit, the lepton-lepton 2NEP takes the form \cite{LeThien}
\begin{equation}
V_{\alpha \beta}(r)  \simeq  \frac{G_{F}^{2}}{4\pi^{3}r^{5}}\left[\sum_{a = 1}^{3} (g_{V}^{\alpha} +|U_{\alpha a}|^{2} ) (g_{V}^{\beta} +|U_{\beta a}|^{2} ) + 2 \sum_{a>b}^{3} \Re( U_{\alpha a}^* U_{\alpha b}^{}  U_{\beta b}^* U_{\beta a}^{} )\right].
\label{V alpha beta massless limit}
\end{equation}
There are four main important remarks about the short-range lepton-lepton 2NEP in Eq.~(\ref{V alpha beta massless limit}). Firstly, the dependence on the mixing parameters drops out at short distances for the 2NEP between same generation leptons. The explanation for this behavior is similar to the one in Section~\ref{nucleon lepton subsection}. Secondly, the lepton-lepton 2NEP exhibit dependence of the mixing parameters across all distances. While this is obvious at long distances based on Eq.~(\ref{E vac 2 lepton lepton massive aa})-(\ref{E vac 2 lepton lepton massless}), the explanation at short distances is not so trivial. Because through the CC interaction leptons only emit and absorb their corresponding neutrino flavors, in order for an $\alpha$-neutrino (antinenutrino) emitted from an $\alpha$-lepton  to be absorbed by a $\beta$-lepton, the $\alpha$-neutrino (antineutrino) must always undergo oscillation into a $\beta$-neutrino (antineutrino). Hence, regardless how short the distance between two leptons from different generation is, the neutrino flavor fields between them must always oscillate, resulting in the mixing-parameters dependence. Thirdly, the Dirac CP-violation phase only manifests at large distances, the reason for this is not well-understood at the moment. Lastly, $V_{ee}(r)$ does not depend on the Dirac CP-violation phase across all distances.           

\section{Discussion}

We derived the nucleon-nucleon, nucleon-lepton and lepton-lepton 2NEP with the phenomenology of neutrino mixing incorporated for Dirac neutrinos. It is demonstrated that due to the virtual exchange process, many fundamental properties can be studied via the 2NEP. Furthermore, by developing a new approach to calculate the 2NEP we also show explicitly the sensitivity of the 2NEP to the structure of the neutrino free vacuum, while current neutrino experiments are not since they only have access to ultrarelativistic neutrinos. Lastly, our work also shows that the effect of neutrino mixing on the 2NEP here is due to the bilinear structure of the SM weak interaction in the low-energy limit.

From the results in Section~\ref{result}, it is clear that the 2NEP is a promising arena to study many fundamental properties of neutrinos. While this force is certainly minuscule, there is evidence to suggest that it is very close to experimental reach.\cite{Stadnik,fischbach et al} Generally, there are 2 main approaches which can be categorized in terms of distance scales. In the short-range limit $\lesssim$nm (assuming the heaviest neutrino $\sim$eV), within the SM, Eq.~(\ref{V alpha beta massless limit}) may offer an alternative measurement of the mixing parameters except for the Dirac CP-phase violation. A good experimental candidate for this approach is to do precision spectroscopy with muonium. Based on our results, an estimate for the effect from the 2NEP is $\sim 1$ Hz, while the current experimental error for this system is $\sim 53$ Hz.\cite{Frugiuele} In the long-range limit $\gtrsim$$\mu$m, precision measurement of macroscopic forces provide a direct measurement for the neutrino masses. Our result indicates that for the range of the lightest neutrino mass $\ll 0.001$ eV, the 2NEP is $10$ to $15$ orders of magnitude away from current submillimeter mechanical force measurements.\cite{Force PRL}  While at present the direct detection of the 2NEP at these distances may seem daunting, the development of new techniques in the measurement of short-range forces and the boundless ingenuity of experimentalists make one optimistic that the 2NEP will be observed in nature someday.

\section*{Acknowledgments}

We thank Ephraim Fischbach for useful conversations and earlier papers on the 2-NEP, which provided significant motivation for our work. We also thank Sheakha Aldaihan and Mike Snow for discussions on the derivation of potentials, which influenced our approach. Q. Le Thien would like to express his deepest gratitude for Professor J. Tran Thanh Van and the ICISE for the hospitality and the generous financial support, which enabled him to present our work at this conference.


\section*{References}

\begin{thebibliography}{99}


\bibitem{Stadnik}  Y.V.\ Stadnik, Phys. Rev. Lett. {\bf 120}, 223202 (2018).

\bibitem{Asaka}  T.\ Asaka, M.\ Tanaka, K.\ Tsumura, and M.\ Yoshimua, arXiv:1810.05429.

\bibitem{LeThien} Q.\ Le Thien and D.E.\ Krause, Phys.\ Rev.\ D {\bf 99}, 116006 (2019).

\bibitem{feinberg}G. Feinberg and J. Sucher, Phys. Rev. {\bf 166}, 1638 (1968).

\bibitem{fischbach et al}E. Fischbach, D. E. Krause, C. Talmadge, and D. Tadić, Phys. Rev. D {\bf 52}, 5417 (1995).

\bibitem{Fischbach AoP} E.\ Fischbach, Ann.\ Phys. (NY) {\bf 247}, 213 (1996).

\bibitem{choudhury} S. Roy Choudhury and S. Choubey, JCAP {\bf 09}, 017 (2018).

\bibitem{BV AoP} M. Blasone, G. Vitiello, Ann. Phys. (NY) {\bf 244}, 283 (1995).

\bibitem{Tureanu} A. Tureanu, arXiv:1902.01232.

\bibitem{Shrock PLB} R. E. Shrock, Phys. Lett. B {\bf 96}, 159 (1980). 

\bibitem{Shrock PRD} R. E. Shrock, Phys. Rev. D {\bf 24}, 1275  (1981).

\bibitem{Ho}  C. M. Ho, J. High Energy Phys. 12,  (2012) 022.

\bibitem{Frugiuele} C. Frugiuele, J. P\`erez-R\`ios, C. Peset, Phys. Rev. D {\bf 100}, 015010 (2019).

\bibitem{Force PRL} W.-H. Tan, S.-Q. Yang, C.-G. Shao, J. Li, A.-B. Du, B.-F. Zhan, Q.-L. Wang, P.-S. Luo, L.-C. Tu, and J. Luo, Phys. Rev. Lett. {\bf 116}, 131101 (2016).

\end{thebibliography}
\end{document}